\documentclass{article}
\usepackage[final]{neurips_2019_ml4ps}

\usepackage{amsmath,amssymb}
\usepackage{hyperref}
\usepackage[dvipsnames]{xcolor}
\usepackage{cuted} %
\usepackage{subcaption}
\usepackage{graphicx}

\usepackage{comment}

\usepackage{float}
\allowdisplaybreaks

\title{Deep convolutional neural networks for multi-scale time-series classification and application to disruption prediction in fusion devices}

\author{R.M. Churchill \\
Theory Department \\
Princeton Plasma Physics Laboratory \\
100 Stellarator Road, Princeton, NJ 08540, USA \\
\texttt{rchurchi@pppl.gov}
 \And
 and the DIII-D team \\
 General Atomics \\
 P.O. Box 85608, San Diego, California 92186, USA}

\begin{document}

\maketitle

\begin{abstract}
\hspace{1cm} The multi-scale, mutli-physics nature of fusion plasmas makes predicting plasma events challenging. Recent advances in deep convolutional neural network architectures (CNN) utilizing dilated convolutions enable accurate predictions on sequences which have long-range, multi-scale characteristics, such as the time-series generated by diagnostic instruments observing fusion plasmas. Here we apply this neural network architecture to the popular problem of disruption prediction in fusion tokamaks, utilizing raw data from a single diagnostic, the Electron Cyclotron Emission imaging (ECEi) diagnostic from the DIII-D tokamak. ECEi measures a fundamental plasma quantity (electron temperature) with high temporal resolution over the entire plasma discharge, making it sensitive to a number of potential pre-disruptions markers with different temporal and spatial scales. Promising, initial disruption prediction results are obtained training a deep CNN with large receptive field ({$\sim$}30k), achieving an $F_1$-score of {$\sim$}91\% on individual time-slices using only the ECEi data.
\end{abstract}

\section{\label{sec:intro}Introduction}
\begin{figure}
    \centering
    \includegraphics[width=0.8\linewidth]{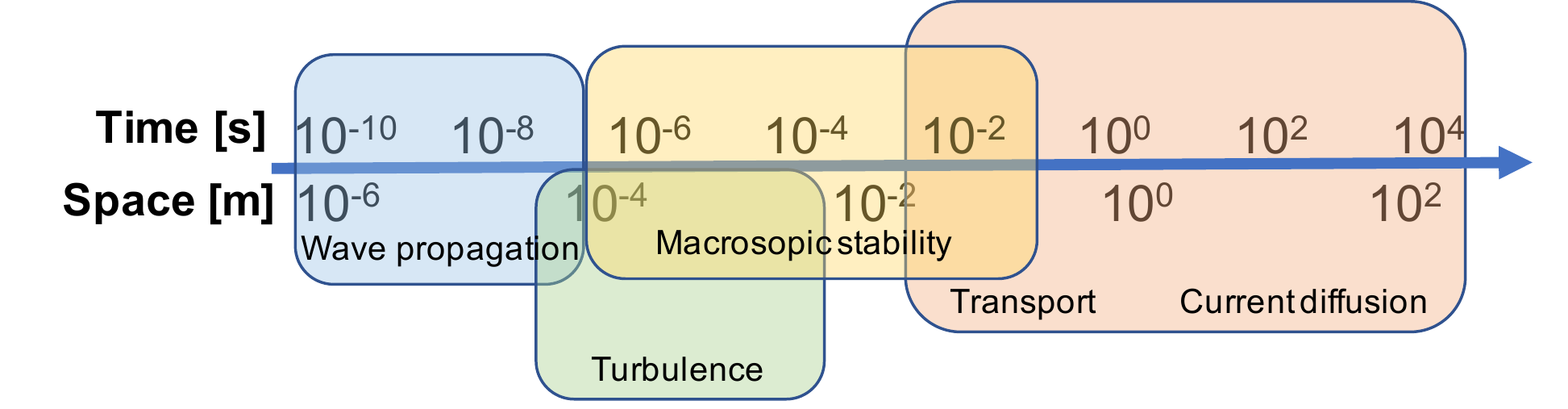}
    \caption{Example temporal and spatial scales of different broad physics phenomena in fusion plasmas, based on Ref.  \cite{fsp2002}}
    \label{fig:scales}
\end{figure}

Plasma phenomena contain a wide range of temporal and spatial scales, often exhibiting multi-scale characteristics (see Figure \ref{fig:scales}). In fusion energy plasmas, many disparate diagnostic instruments are simultaneously used in order to capture these various spatiotemporal scales, and to cover the multiple physics present in these plasmas. In addition, fusion experiments are increasingly built to run longer pulses, with a goal of eventually running a reactor continuously. The confluence of these facts leads to large, complex datasets with phenomena manifest over long sequences. A key challenge is enabling scientists/engineers to utilize these long sequence datasets to, for example, automatically catalog events of interest or predict the onset of phenomena.

Many deep learning architectures have been created and successfully applied to sequence learning \cite{Graves2013, LeCun2015, Lipton2015,Fawaz2018} problems, in areas of time-series analysis or natural language processing. However, many of the typical architectures used for learning from sequences (e.g. recurrent neural networks (RNN) and its most popular variant Long Short Time Memory networks (LSTM))  suffer from memory loss; long-range dependencies in sequences are difficult for these architectures to track \cite{Bai2018}.

In this paper we discuss recent advances in neural networks, specifically an architecture that uses dilated convolutions in a deep convolutional neural network (CNN), which was designed to overcome these problems of learning on long sequences. We use this architecture to predict oncoming disruptions in fusion plasma discharges of the DIII-D tokamak utilizing only raw data from a single, high temporal resolution imaging diagnostic (the Electron Cyclotron Emission imaging diagnostic, or ECEi) \cite{Tobias2010}. Because the ECEi diagnostic is sensitive to a range of multi-scale dynamics in the plasma related to disruptions \cite{Choi2016}, it offers the potential to more accurately predict them. Avoiding disruptions is a grand challenge for tokamak fusion devices on the road to fusion energy \cite{Hender2007}. While much research has gone into utilizing machine learning for disruption prediction \cite{Vega2013,Rea2018,Kates-Harbeck2019}, often global, reduced 0-D features are used in shallow machine learning methods. Recently work utilizing deep LSTM networks also added the use of low temporal resolution 1-D plasma profiles \cite{Kates-Harbeck2019}, and another work used a combination CNN/LSTM on resampled, low temporal resolution bolometer data \cite{Ferreira2018}. The work we present here takes inspiration from these works in utilizing higher dimensional signals, and shows how to use newer deep learning architectures to learn on high-temporal resolution data with long-range dependencies due to multi-scale physics.  

\section{\label{sec:cnn_dilated}Deep convolutional neural networks with dilated convolutions}
Recently there has been much research into deep learning architectures which can overcome the deficiencies of RNN/LSTM's, and handle long, multi-scale sequences \cite{Vaswani2017,Devlin2018,Radford2019,Wang2017,Gehring2017}. A seminal paper presented one such architecture, WaveNET \cite{VanDenOord2016}, which is a convolutional neural network (CNN) focused on generating realistic audio. One of the key insights of this paper was to use dilated convolutions to increase the receptive field of the network. This overcomes the dilemma faced with using normal convolutions in causal networks, where to be sensitive to long sequences you must increase the convolutional filter size and/or the number of layers in the network. Dilated convolutions have a dilation factor ($d$) which represents the number of input points skipped between filter parameters, e.g. the sequence output $y[n]$ from a dilated convolution with dilation $d$ is:
$$
y[n] = \sum_{i=0}^{k-1} w[i] \, x[n-d\cdot i] 
$$
where $w$ represents the weights of the 1D dilated convolution filter of length $k$, and $x[n]$ is the input sequence. A normal convolution results by setting $d=1$. By stacking layers of dilated convolutions, and increasing the dilation factor in each layer, the receptive field of the network can be increased while maintaining a tractable number of model parameters. 

Dilated convolutions impose an inductive bias or specific structure to the architecture which guide the transformations learned by the neural network. Specifically, dilated convolutions have a natural connection with wavelet structures, which have been used for separating out structure in multi-scale data, including turbulent flows \cite{Farge1992}. In a loose sense, these neural networks allow us to learn the wavelet coefficients needed to accomplish our classification task.

A simplified yet powerful architecture named temporal convolutional network (TCN) \cite{Bai2018} built upon this WaveNET work, utilizing dilated convolutions and many modern neural network techniques, such as weight normalization and residual connections.  Bai \emph{et. al.} \cite{Bai2018} showed the TCN could outperform LSTM and GRU architectures on many common sequence learning tasks, especially for long sequences with long-range dependencies. It is this TCN architecture that we will now apply to the problem of disruption prediction using ECEi data.

\section{\label{sec:apply}Application to Disruption Prediction using Raw ECEi Imaging Data}
Disruptions in tokamaks plasmas are a sudden loss of control which cause a termination of the plasma and potentially large destructive forces and/or heating on the containment vessel and protective wall materials. Next-step devices such as ITER and beyond will have a low tolerance for disruptions \cite{DeVries2016}. We need to ensure disruptions can be avoided by accurate prediction of oncoming disruptions and mitigation techniques if necessary. 

Here we apply the TCN architecture to high-temporal resolution, raw ECEi imaging data from the DIII-D tokamak for the purpose of predicting oncoming disruptions\footnote{Code available at \url{https://github.com/rmchurch/disruptcnn}}. 

\subsection{Data}
\begin{figure}
    \centering
    \includegraphics[width=0.8\linewidth]{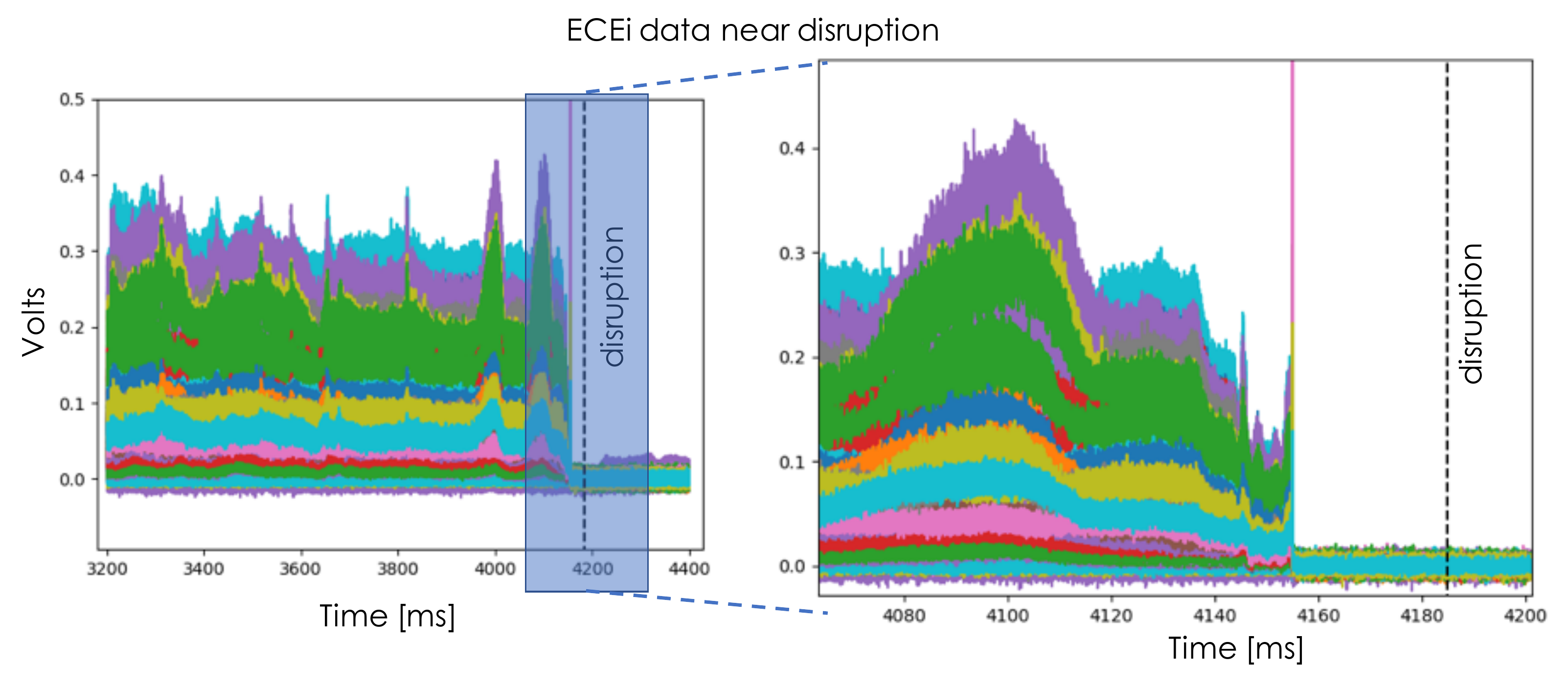}
    \caption{DIII-D ECEi diagnostic time-series data from each of the 160 channels of the LFS diagnostic, near a disruption event. The sudden drop in ECEi signal a few milliseconds before the disruption time (which is set at the current quench time) is due to the drop in temperature at the thermal quench.}
    \label{fig:ecei}
\end{figure}

The ECEi diagnostic \cite{Tobias2010} is used to measure electron temperature on very fast timescales, normally sampling at 1 MHz on the DIII-D tokamak. The diagnostic has 160 spatial channels, laid out in a rectangular grid with 20 vertical by 8 radial channels. Example time series of the DIII-D ECEi diagnostic near a disruption is shown in Figure \ref{fig:ecei}.  ECEi can capture a number of plasma phenomena such as turbulence fluctuations, tearing modes, sawteeth, and ELMs \cite{Tobias2010}, which allow it to be sensitive to a number of pre-disruption markers. A dataset of good ECEi data (SNR$>3$) from 2,747 DIII-D shots ( ${\sim}42\%$ disruptive, ${\sim}58\%$ non-disruptive) was selected, measuring about 10 TB. Time length of each shot varies, typically between 5 to 10 seconds. Raw digitizer voltage output was corrected for digitizer drift, then $z$-normalized before inputing into the TCN. For ease of training the neural network, we decided as an initial step to temporally downsample the ECEi data to 100 kHz (i.e. factor of 10x less data). 

\subsection{Model and Training Setup}
We treat the problem of disruption prediction as a binary classification problem, where we predict whether each time slice corresponds to a ``non-disruptive'' or ``disruptive'' class. We label all time slices within 300ms of a disruption as ``disruptive'' ($t_{disrupt}-t<300$ms), and all other time slices as ``non-disruptive'' \cite{Rea2018} (sequences from shots without disruptions are taken during established times of the discharge, i.e. during the plasma current flattop). Typical binary cross-entropy loss is used as the loss function for the neural network training. 

We define our TCN model to have a receptive field of $N_{recept}{\sim}30,000$. This is an order of magnitude larger than receptive fields in the original TCN \cite{Bai2018} or WaveNET \cite{VanDenOord2016} papers. With the 100 kHz sampling rate, this means that each time slice prediction uses the previous $\sim 300$ms in order to make the prediction. With our definition of disruptive time slices as within 300ms of the disruption, this implicitly assumes that 600ms before a disruption is sufficient to predict oncoming disruptions. We use a 4 hidden layer TCN with dilations $[1,10,100,961]$ (i.e. increasing by a factor of about 10 each layer), with a filter kernel size of 15. The number of filters per hidden layer was held constant at 80 (varying number of filters per hidden layer was not attempted).

The TCN architecture allows parallelization of the sequence prediction by inputting sequences of length $N_{seq}$, which are longer than $N_{recept}$, resulting in $N_{seq}-N_{recept}+1$ predictions per sequence. Empirically it was found that sequence lengths of $N_{seq}=78,125$ allowed model computations that fit inside the GPU memory constraints, while allowing a batch size of 12 (per GPU) to ensure sufficient variety within each batch for training with stochastic gradient descent (the total batch size with data parallelism was 192. Larger batch size can be achieved reducing the sequence length, though at an increased computational cost due to more data reads). The set of sequences with timeslices consisting of only the majority class (``non-disruptive'') was undersampled such that there were balanced disruptive and non-disruptive sequences.

Stochastic Gradient Descent (SGD) with Nesterov momentum 0.9 was used to train the model, with an initial learning rate of 0.5 that was decreased automatically upon plateau (\texttt{ReduceLROnPlateau}). A warmup period was used for the first 5 epochs, increasing the learning rate from 0.0625 to 0.5 to enable larger batch training \cite{Goyal2017}. Multi-node, multi-GPU setup was used to parallelize the training. The Pytorch built-in synchronous data parallel training routine \texttt{DistributedDataParallel} was used \cite{Paszke2017}, training on 16 GPUs over 2 days.

\subsection{Results}
\begin{figure}
    \centering
    \begin{subfigure}{0.48\textwidth}
    \includegraphics[width=\textwidth]{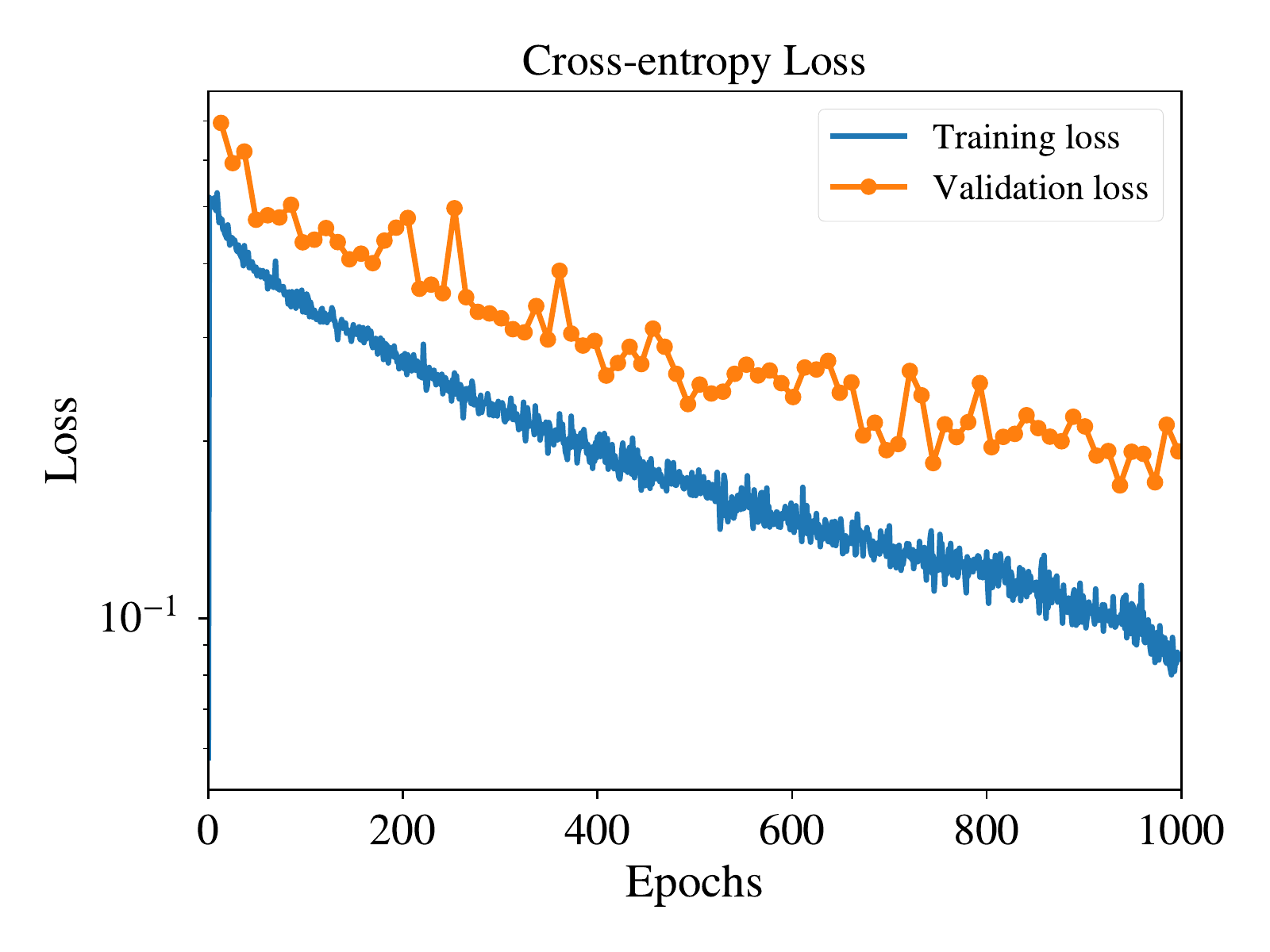}
    \label{fig:resultsa}
    \end{subfigure}
    \begin{subfigure}{0.48\textwidth}
    \includegraphics[width=\textwidth]{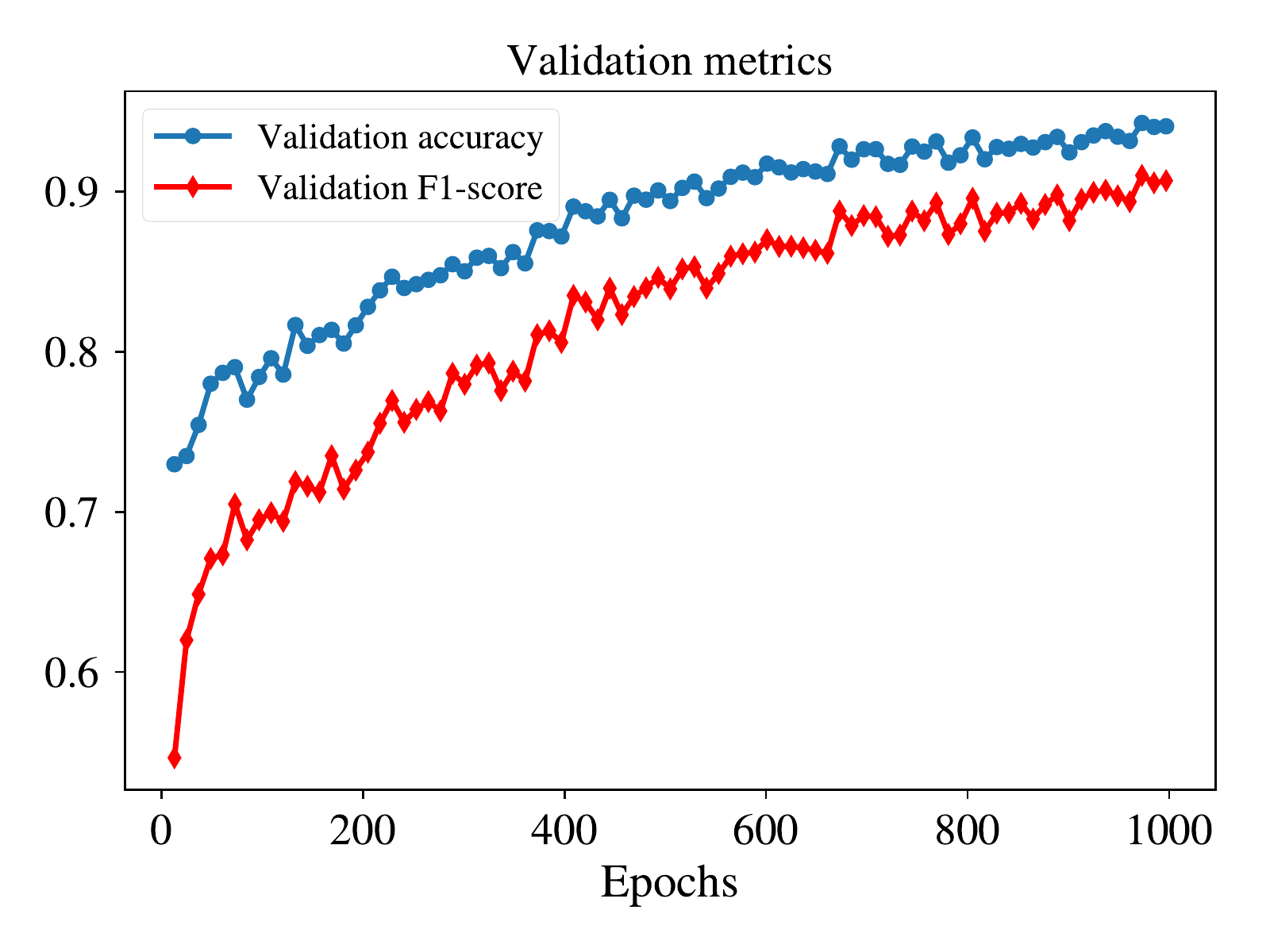}
    \label{fig:resultsb}
    \end{subfigure}
    \caption{Results from training the TCN on ECEi data. }
    \label{fig:results}
\end{figure}

The results of training this TCN model on ECEi data for disruption prediction on DIII-D are shown in Figure \ref{fig:results}. Results are plotted over 1000 training epochs. The training binary cross-entropy loss continually decreases over the training, showing our model has the capacity to learn the task from this dataset. The validation loss also continually decreases, slightly flattening towards the end, indicating the model is reaching the limit of its generalizability after 1000 epochs. Two validation metrics are also shown: accuracy (how many time slices were predicted correctly as disruptive or non-disruptive), and F1-score (a geometric mean between precision and recall). Because the time slice classes are imbalanced (even though the sequence sets are balanced), the F1-score gives a better indication of how well our classifier does on the minority class (disruptive).

The metric of accuracy reaches ${\sim}94\%$, but more importantly the metric of F1-score reaches ${\sim}91\%$, showing the neural network has learned to predict individual time slices of both disruptive and non-disruptive time slices very well. Current machine learning disruption predictors typically achieve a true-positive rate in the low 90\% on shots \cite{Vega2013,Rea2018,Kates-Harbeck2019,Ferreira2018}, with the goal of ${>}95\%$ with a false-positive rate of ${<}5\%$\cite{DeVries2016}. The results presented here offer a promising path to overcome this gap. Consolidation of the time slice predictions to make shot predictions is left for future work, including not training on the last 30 ms before a disruption since this is a minimum amount of time needed to trigger mitigation systems. Most likely techniques like the hysteresis threshold algorithm will be needed \cite{Montes2019}, as the fast, noisy ECEi data could easily trigger occasional false predictions.

\section{Discussion and Future Work\label{sec:future}}
These results show the usefulness of deep convolutional neural networks with dilated convolutions for fusion problems where the multi-scale, multi-physics nature mandates capturing long-range dependencies in time-series. They have shown that it is possible to apply deep learning directly on the raw data from a single diagnostic with high temporal resolution in order to make useful disruption predictions, a topic critical to the success of magnetic confinement fusion. They also show that training TCN networks with large receptive fields on the order of ${\sim}30k$ is possible, allowing learning on long sequences with long-range dependencies. 

Future work in various areas is planned. At the base level, using the full dataset at full temporal resolution could give further improvement, though may require model parallelism to train. Further, combining multiple modalities (including more diagnostics) \cite{Dumoulin2018} can allow the disruption predictions to be sensitive to the various physics which can trigger disruptions \cite{DeVries2016}. Also, interpretability of the network decisions is highly desired, especially to understand the physics and extend to future machines \cite{Kim2017}.

\subsubsection*{Acknowledgments}
The main author would like to thank Ben Tobias, Yilun Zhu, Dave Schissel, C.S. Chang, Bill Tang, Julien Kates-Harbeck, Raffi Nazikian, Cristina Rea, Bob Granetz, Neville Luhmann, Sean Flanagan, Ahmed Diallo, and Ken Silber for various contributions to this work. This material is based upon work supported by the U.S. Department of Energy, Office of Science, Office of Fusion Energy Sciences,  under AC02-09CH11466, DE-FC02-04ER54698, and FG02-99ER54531. We also recognize the Princeton Research Computing center for the computational resources used in this paper.

This report was prepared as an account of work sponsored by an agency of the United States Government.  Neither the United States Government nor any agency thereof, nor any of their employees, makes any warranty, express or implied, or assumes any legal liability or responsibility for the accuracy, completeness, or usefulness of any information, apparatus, product, or process disclosed, or represents that its use would not infringe privately owned rights. Reference herein to any specific commercial product, process, or service by trade name, trademark, manufacturer, or otherwise, does not necessarily constitute or imply its endorsement, recommendation, or favoring by the United States Government or any agency thereof. The views and opinions of authors expressed herein do not necessarily state or reflect those of the United States Government or any agency thereof.

\bibliographystyle{abbrv}
\bibliography{neurips2019_final_arxiv}

\end{document}